# Spatiotemporal evolution of a self-excited dust density wave in a nanodusty plasma under strong Havnes effect

Bidyut Chutia[1], T. Deka[1,2], Y. Bailung[1,3], D. Sharma[1,4], S. K. Sharma[1,5*] and H. Bailung[1]

[1]*Physical Sciences Division, Institute of Advanced Study in Science and Technology (IASST), Paschim Boragaon, Guwahati-781035, Assam, India*

[2]Presently at *Department of Physics, Sipajhar College, Mangaldoi-784145, Assam, India*

[3]Presently at *Department of Physics, Goalpara College, Goalpara-783101, Assam, India*

[4]*Academy of Scientific and Innovative Research (AcSIR), Ghaziabad-201002, India*

[5]Presently at *Department of Physics, Assam Don Bosco University, Tapesia Gardens, Sonapur-782402, Assam, India*

*sumita_sharma82@yahoo.com

A broad-spectrum self-excited dust density wave is experimentally studied in a vertically extended nanodusty plasma consisting of *in situ* grown carbonaceous nanometer sized particles. The nanodusty plasma having high particle density (of the order of $10^{12} - 10^{13}\ m^{-3}$) is created with vertical extension up to $(40 \pm 0.1)\ cm$ and radial extension up to $(5 \pm 0.1)\ cm$. The propagation of the self-excited dust density wave under strong Havnes effect is examined over a large axial distance $(19 \pm 0.1)\ cm$. Time-resolved Hilbert transformation and Fast Fourier transformation techniques are used to study the spatiotemporal evolution of frequency and wave numbers along three directions from the dust void viz. axial, radial and oblique. The propagation is found to be inhomogeneous throughout the dust cloud. The phase velocity of the wave is estimated to be quite low and decreasing along the direction of propagation. This effect is attributed to the strong reduction of particle charge due to a high Havnes parameter along the propagation direction. By the estimation of average particle charge, ion density and the finite electric field throughout the nanodust cloud, a quantitative analysis of the void formation in nanodusty plasma is presented. New insights are also made regarding wave merging phenomena using time-resolved Hilbert transformation.

# I. INTRODUCTION

Dusty (complex) plasma i.e., electron-ion plasma that contains solid particulates of size varying from a few nanometer (nanodusty) to micrometer (micron dusty) are omnipresent in astrophysical as well as in laboratory environments [1,2]. In laboratory, the dust particles get charged up negatively by the collection of electrons and ions from the background plasma. The heavily charged dust particles then get confined in the sheath region (micron sized particles) or in the bulk plasma itself (nanometer sized particles). Being electrically charged, the dust particles become an active component of plasma which changes the entire dynamics of the system and supports a new range of phenomena [3–6]. Dusty plasmas are enriched with a large number of physical phenomena such as waves and instabilities [7], lattice waves [8–10] or phonons [11,12], strong-coupling effects [13–15], non-linear solitary and shock formations [16–22], dust voids [23–25], vortices [26–32] etc. Among such vast range of phenomena, dust density (acoustic) wave has been a dynamic topic of research in dusty plasma physics since its theoretical prediction by Rao *et al.* [33] in 1990 and experimental confirmation by Chu *et al.* [34] and Barkan *et al.* [35] in 1995. Dust density waves (DDWs) are compressional waves analogous to the ion acoustic waves where massive dust particles provide the inertia. Whereas, the electric field produced due to the charge separation during compression/rarefaction of the dust particles provides the restoring force. Being larger in size, the dust particles retain a large negative value of electric charges and on the other hand the massiveness of the particles provides a unique opportunity to visualize their low frequency (~ Hz) dynamics using high-speed cameras [36,37]. Because of these unique features of dusty plasma one can study various phenomena at their most kinetic level.

In laboratory, DDWs have been excited both externally [38] or spontaneously (self-excited) [35,39,40]. Self-excited DDWs arise due to the ion-streaming instability and propagate along the direction of ion streaming. In some experiments, these DDWs are also found to be propagating obliquely to the direction of ion streaming [41]. Self-excited DDWs are very common throughout the bulk of nanodusty plasmas (as the nanoparticles have low mass compared to micron sized particles) [42]. For dusty plasma with micron sized grains such waves are investigated in microgravity condition using parabolic flights and in International Space Station [43–46]. In normal laboratory condition such self-excited waves are observed in the anodic discharge zone where the ion streaming can be strong enough to drive the wave [47,48]. In the last 25 years, several theoretical and experimental studies have been performed to explore linear and non-linear characteristics of the DDW mostly in micron dusty plasmas. A significant number of studies have been dedicated to explore wave-particle interaction [38,49,50], dispersion relation [47,51–55], wave breaking or merging [56,57], shocks [18,58], synchronization [47,51–53,55,59–62], existence of frequency clusters [60,63,64] and turbulence [65,66]. Recently, the effect of magnetic field (from a few millitesla a few hundred millitesla) on the propagation of dust density wave is studied in detail [67–69]. The wave activity is found to cease at higher magnetic field.

In case of nanodusty plasmas many new effects come into the picture as the dense particle cloud strongly influences the background plasma as well as the discharge characteristics. A strong reduction in plasma electrons in highly dense nanodusty plasma is one of the main characteristics which in turn enhances the Havnes effect [39,70]. The Havnes parameter is proportional to the ratio of dust density to the ion density of the plasma and becomes greater than unity for higher dust density. Under this circumstance the charge neutrality condition in the plasma is drastically modified. Recently, *Tadsen* et al. [39] and *Kortshagen* et al. [71] have shown the use of DDW characteristics to investigate nanodusty plasmas to estimate dust particle charge as well as plasma parameters inside a dense nanodust cloud. Although nanodusty plasmas provide great scopes to study DDW propagation, very less works has been reported compared to micron dusty plasmas.

In this work, the effect of strong electron depletion (Havnes effect) on the propagation of DDW through an extended volume of nanodusty plasma cloud has been investigated. The wave dynamics have been studied along three directions (axial, radial and oblique) from the dust void in a 2D central plane of nanodusty plasma. It is to be noted that the cloud is vertically extended to a very large distance which provided a great scope to study the DDW dynamics over many wavelengths. Standard techniques such as periodograms, time-resolved Hilbert transformation, Fast Fourier transformation etc. are used to characterize the DDW. Wave spectra and dispersion relations are determined and compared with a theoretical model from which the plasma parameters inside the dust cloud are obtained in the 2D central plane. The estimated plasma parameters obtained by using the DDW analysis along the three directions provided a quantitative understanding of the formation of the dust void which is commonly observed in most complex plasma systems with rf discharge. The collision/merging events between two high amplitude wave fronts have also been observed and analysed using the time-resolved Hilbert transformation. Section II describes the experimental setup and procedure, Section III and IV present various experimental results and discussion of the obtained results, respectively. Finally, conclusion is made in section V.

## II. EXPERIMENTAL SETUP AND PROCEDURE

The experiment is performed in the nDuPlEx (nano Dusty Plasma Experiments) setup, which consists of a vertically mounted quartz tube of length 50 cm and inner (outer) diameter 5.5 (6.0) cm. The schematic of the experimental device is shown in Fig. 1. The same experimental setup described elsewhere is used [72]. To produce the Argon (Ar) plasma, a capacitively coupled rf discharge (1-5 W, 13.56 MHz) is applied to the live electrode at the central position on the outer wall of the tube. A stainless-steel (SS) grid is used as a grounded electrode inside the tube placed at the bottom. Both the top and bottom SS flanges are electrically grounded. The flow of Ar, in turn the neutral pressure, is controlled by a mass flow controller (MFC) at a rate of (2-10) sccm. The pressure for this flow rate regime is (9.5-23) Pa. The plasma properties in Ar plasma (electron temperature, ion density) are measured using a cylindrical Langmuir probe inserted through the top flange as shown in Fig. 1. The nanodusty plasma medium is achieved by growing carbonaceous (a-C:H) nanoparticles from gas phase polymerization using reactive discharge of Ar-$C_2H_2$. The rf power used to grow the nanoparticles is 1 W and the flow rate of $C_2H_2$ is maintained at 2 sccm. The total pressure is 20 Pa during Ar-$C_2H_2$ discharge.

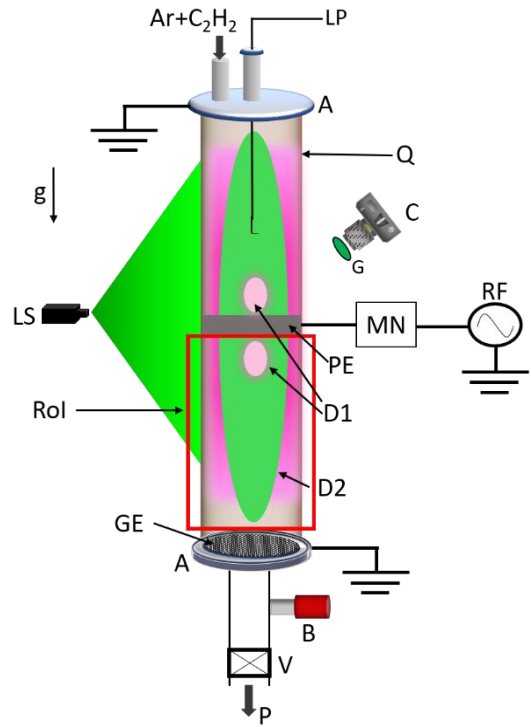

FIG. 1. Schematic diagram of the experimental setup. LP: Langmuir Probe, A: Stainless-Steel Flange, Q: Quartz Tube, G: Green filter, C: Camera, MN: Matching Network, RF: Radio Frequency, PE: Powered Electrode, D1: Dust Void, D2: Dust Cloud, B: Baratron capacitance manometer, V: Angle Valve, P: Rotary Pump, GE: Grid (grounded) Electrode, RoI: Region of Interest, LS: Laser Source, g: acceleration due to gravity.

Such kind of particle growth phenomena are very common in reactive plasma environment containing silane (SiH$_4$), hydrocarbon gases (C$_2$H$_2$, CH$_4$) and chlorofluorocarbons [73]. The growth phenomenon in reactive plasma environment can be classified into three main successive stages such as nucleation, coagulation and accretion [73–75]. Firstly, the reactive gas such as the C$_2$H$_2$ dissociates in the plasma environment and forms the precursors for gas phase polymerization of clusters with a size of a few nanometers. The electric charges of these clusters may fluctuate between a few positive to negative elementary charges along with some neutrals. As soon as the clusters reach a critical density, coagulation of these clusters gets initiated and as a result solid particulate of a few tens of nanometers size are formed. A sudden drop in the cluster density has been observed in this stage. These particulates then collect electrons from the plasma and get charged negatively. The mutual electrostatic repulsion then stops further growth due to coagulation and particulates achieve a linear growth due to surface collection of molecules from the reactive plasma. This stage is called as the accretion. After this state the particles may grow further if C$_2$H$_2$ flow is kept on and ultimately gets pushed out from the plasma confinement due to the force balance and a new growth cycle starts to appear. If the source of C$_2$H$_2$ is stopped in between then the particle size reaches a saturation and stay confined within the bulk plasma as long as desired, with a little to no loss. The detail characteristics of the growth kinetics of the nanoparticles in the system for the present discharge conditions in the nDuPlEx device can be found in our previous work [72]. At 10 min, the C$_2$H$_2$ flow is stopped and a nanoparticle cloud filling the whole chamber has been obtained with a narrow particle size distribution having average radius of 308 nm. The size distribution of the nanoparticles is obtained by collecting the particle using a specially designed particle collector and then by performing ex-situ diagnostic technique such as SEM. To observe the dynamics of the cloud, the central vertical cross-section of the 3D dust cloud is illuminated with a 2D green laser sheet (532 nm, 100 mW) having a constant intensity along the line of expansion. A dust density wave is also seen to appear near the dust void edge which propagates vertically downward to the bottom electrode (along $g$). The device is useful to generate a very long 3D nanodusty plasma column to follow the DDW over a distance of $\sim 19\ cm$. A high-speed camera (Phantom Miro M110 @400-1000 fps) is used to record the optical signals of the DDW. A band pass filter is used to capture the light scattered by the dust cloud only and neglect the light emitted by the plasma.

## III. EXPERIMENTAL RESULTS
### A. Dust density distribution of the nanodust cloud

The spatial distribution of dust number density along the vertical axis is obtained from the pixel intensity of the scattered light by considering a thin rectangular section in the dust cloud. The scattered light intensity is linearly dependent on the dust number density as the camera has a linear response and hence the scattered light intensity can be used marginally to estimate dust density. The intensity profiles for 200 consecutive frames are obtained to get the average intensity profile $I_{av}$ (blue line), which is shown in Fig. 2 (Top panel) as a function of vertical positions. Using this average intensity profile and the known dust densities at 2-3 vertical positions (measured by using laser extinction method as described in [72]), the dust density profile is obtained for the entire selected region and is shown in Fig. 2 (top panel) (red circles).

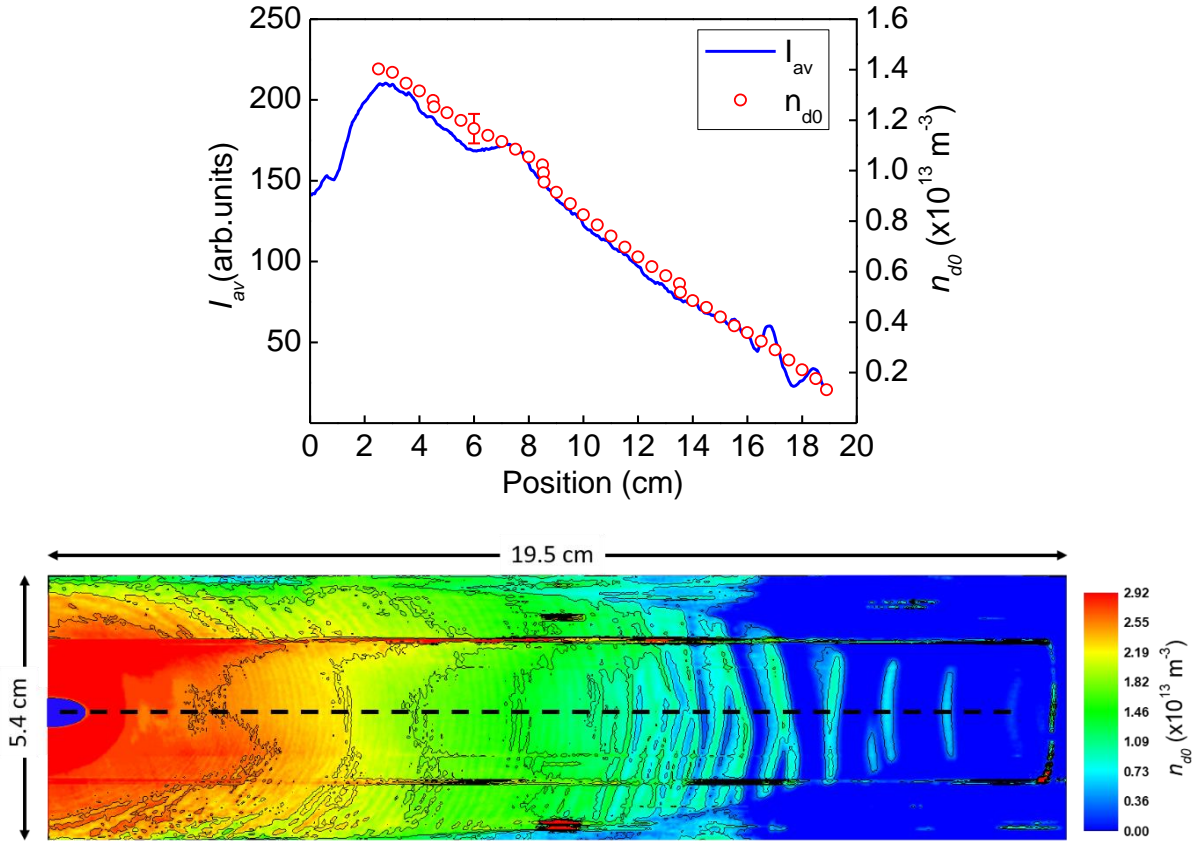

FIG. 2. (Top panel) Average pixel intensity, $I_{av}$ (blue line) and dust density, $n_{d0}$ (red circle) variation as a function of vertical position (along the black dotted line). The zero in the position axis corresponds to the centre of the void. (Bottom panel) Dust density distribution throughout the central slice of the dust cloud. The two straight horizontal lines are the scattered light coming from the quartz tube. The density profile shown in the top figure corresponds to the horizontal dashed line in the bottom figure. The color bar gives the absolute values of dust density.

Using this method, we have computed the average dust density distribution in the entire 2D slice of the cloud by superimposing 200 image frames, which is shown in Fig. 2 (bottom panel). A high gradient of dust density from the void edge towards both axial and radial direction is seen. The highest dust density is around the dust void (of the order of $10^{13}\ m^{-3}$) and gradually decreases (to $10^{12}\ m^{-3}$) near the bottom electrode in the axial direction. However, in radial and oblique direction the dust density is found to be within the order of $10^{13}\ m^{-3}$ because of short extension of the cloud. These quantitative values of dust density are used in calculating the theoretical dispersion relation of the DDW.

### B. Segmented high-speed recording of the DDW

The CMOS high-speed video camera that is used to record the DDW dynamics has a field of view (FoV) of 1280×800 square pixels with a resolution of 20 $\mu m$/pixel. Since the wave field is an extended one (~ 19 cm), the DDW dynamics is recorded from a closer distance by vertically segmenting the RoI (Fig. 1) into three regions, with a slight overlap of the FoV, to have a high spatial resolution. Figure 3 (a), (b) and (c) show the three segmented sections of the whole wave field. These recordings are performed using a single camera by vertically displacing it manually within a few seconds. Each section of the DDW is recorded at a frame rate of 400 fps with an exposure of 0.5 s. It is checked that there is no aliasing effect for our DDW for a frame rate of 400 fps. It has been observed from our previous work [72] that after switching off the $C_2H_2$ flow at 10 min, the particle size, density and the confinement of the cloud remains almost constant. Therefore, it can be assumed that in this condition the wave dynamics also remain same throughout the recording time. To analyse the DDW, periodograms and time-resolved Hilbert transformation are employed in each segmented section of the DDW field. Small regions (yellow rectangles) are selected such as A1, A2 and A3 to perform the analysis axially for the three sections (a), (b) and (c), respectively. In case of (a), additionally two different rectangular regions, B and C are chosen to perform DDW analysis which are shown in Fig. 3 (a). Region As are in the axial direction of the tube, region B is in the radial direction and region C is an oblique one which is at $45^0$ to both region A and B. The effect of coating of the quartz tube due to the reactive discharge of Ar-$C_2H_2$ on the clear recording of DDW is minimal, although we have taken care of that while calculating the dust density, as described in [72].

## C. Periodograms

Periodograms or space-time plots are very effective in understanding the wave propagation as described earlier by various works as it can track single wavefronts over time [38,46,49,63,67,76]. To plot the periodograms, the intensity profiles of the selected rectangular regions for 200 frames (0.5 s) are obtained. These intensity values are then subtracted from the temporally averaged intensity profile for each position which enhance the intensity to a great extent and help us to detect the wave fronts efficiently. Then, periodograms for the three sections A1, A2 and A3 of Fig. 3 are plotted from the averaged intensity values for each spatial position and at each temporal point. These are shown in Fig. 3 (d), (e) and (f), respectively. The dust void edge is shown using a vertical (yellow) dashed line in the periodograms of Fig. 3 (d)-(f). From the periodograms, one can easily track the propagating wave fronts as inclined bright bands from which one can easily find out the frequency, wave number and the phase velocity of the wave (see Fig. SM1 (a) of Supplementary Material). The frequency is obtained from the inverse of the separation between two successive bright bands in the time axis. Similarly, the inverse of the distance between two successive bands along the position axis gives the wave number. Finally, the inverse of slope of the bands yields the phase velocity of the wave. From the periodograms it is seen that the wave fronts are propagating (positive slope) in region A with a large variation of wave frequency and wave number along its direction of propagation (along *g*). The average frequency decreases and the average wavelength increases as the wave propagates throughout the entire length of the cloud. From Fig. 3 (a)-(c), it is seen that the wave fronts are curved near the dust void, where they originate, and as they propagate, they become straight at the bottom of the chamber, in Fig. 3 (c). In Fig. 3 (d), the wave fronts are quite linear (low amplitude) and are coherent but as they propagate (see Fig. 3 (e)), the wave amplitude increases indicating a transition from linear to non-linear regime. In Fig. 3 (f), these wave fronts grow to strongly non-linear and become turbulent. Beyond 7 cm (from the centre of the dust void), the normalized dust density perturbation, $\partial n_d/n_{d0} = (I - I_{av})/I_{av}$, reaches above 20% meaning a non-linear DDW. As the DDW propagation is not very organized it can be noted that the nanodust cloud is probably in a weakly coupled state [76]. The phase velocity also decreases along the direction of propagation as the slope of the bright bands increases in the space-time plot. The merging phenomena of wave fronts are shown in Fig. 3 (f) with dashed squares. It is very clear from the periodograms how the wave fronts merge/collide and form a higher amplitude wave front which propagates at a lower phase velocity as the slope (in the periodograms) is seen to be increased. The points where the wave merging occur are called defects. These defects are the results of existence of frequency clusters in non-linear DDW [60,63]. Such bifurcations of wavefront occur because different regions of a wavefront travels with different velocities. It has also been observed that the wave number is not conserved before and after merging of wavefronts as seen in previous works [46,48,76]. Although the wave is coherent in the central region near the void, the turbulent nature is evident towards the wall of the chamber in all the sections of Fig. 3 (a)-(c). The formation of secondary wave fronts in between two wave crests also can be seen (see Fig. SM1 (b) of Supplementary Material). The study of such secondary wave formation is beyond the scope of this article. In case of region B and C (see Fig. SM2 of Supplementary Material), it can be seen that the wave starts near the void edge with relatively coherent nature and suddenly reach to non-coherent or turbulent regime meaning a building up of non-linearity unlike in region A, where the DDW takes some time to evolve into a turbulent one.

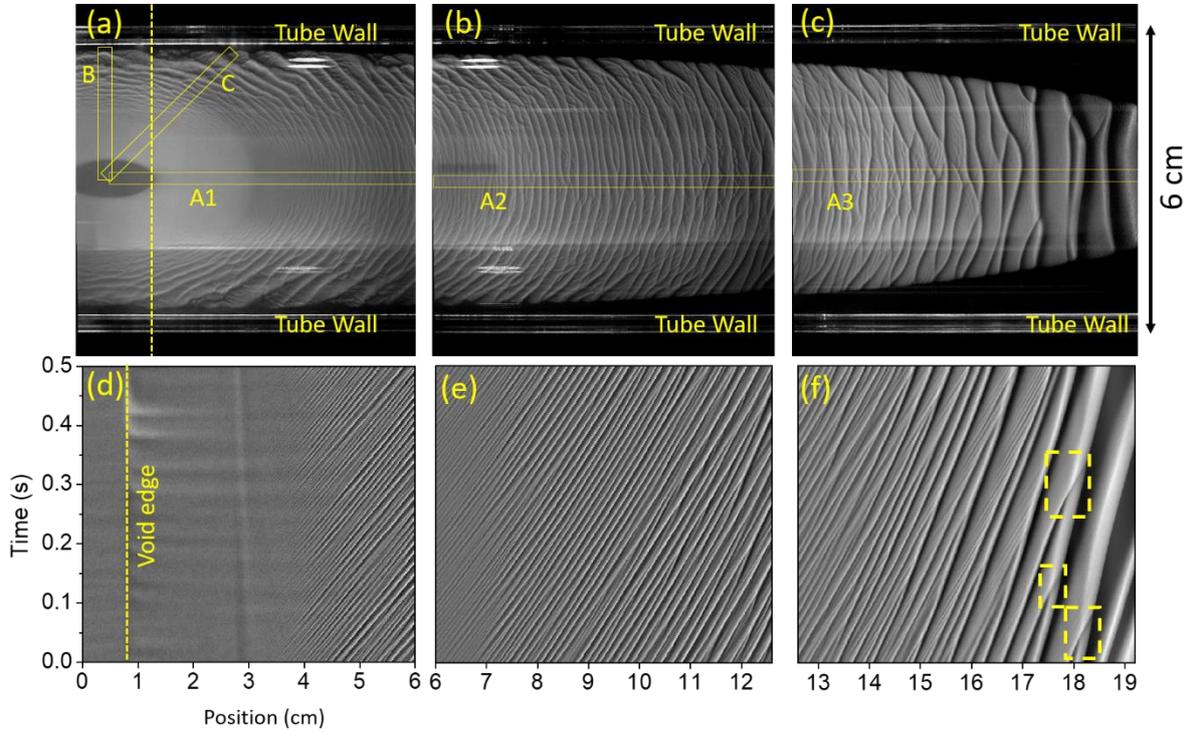

FIG. 3. (a), (b) and (c) are the typical snapshots of the three consecutive segments of the wave field. The thin yellow rectangular boxes A (consists of A1, A2, A3), B and C are the regions considered for wave analysis propagating in three different directions i.e., axial, radial and oblique (at $45^0$ to both A and B), respectively. (d), (e) and (f) are the periodograms of region A1, A2 and A3, respectively. The dashed(yellow) lines in (a) and (d) represent the edge of the void. The zero of the position axes refers to the centre of the dust void. The dashed (yellow) rectangular boxes in (f) represent the wave merging phenomena.

## D. Time-resolved Hilbert transformation

As observed in the periodograms of the DDW, the wave dynamics is quite complex and not homogeneous at all throughout the nanodust cloud. Therefore, a more quantitative analysis of the local properties of wave dynamics is required. Therefore, we employed time-resolved Hilbert transformation [39,46,77,78] to obtain the spatiotemporal evolution wave frequency and wave number by using an analytic signal. The analytic signal is introduced by Gabor [79] in 1946, which is constructed from the time-series of the wave intensity obtained from the recorded video. The time-series signal, $n_d(r,t)$ is obtained by subtracting the time-averaged intensity fluctuation signal, $<n_d(r,t)>$ from the raw intensity signal. The analytic signal $A(r,t)$ is then formulated according to Eq. (1),

$$A(r,t) = n_d(r,t) + i\hat{n}_d(r,t)$$

$$= B(r,t)\exp[i\phi(r,t)], \qquad (1)$$

where $\hat{n}_d(r,t)$ is the Hilbert transform of $n_d(r,t)$, $B(r,t) = (n_d^2 + \hat{n}_d^2)^{\frac{1}{2}}$ is the envelope function, and $\phi(r,t)$ is the instantaneous phase. The next step is to unwrap the instantaneous phase to eliminate any discontinuity due to phase jumps from $2\pi \to 0$. One can obtain the instantaneous frequency, $f(t)$ and wave number, $k$ by performing temporal and spatial derivative of the instantaneous phase, respectively.

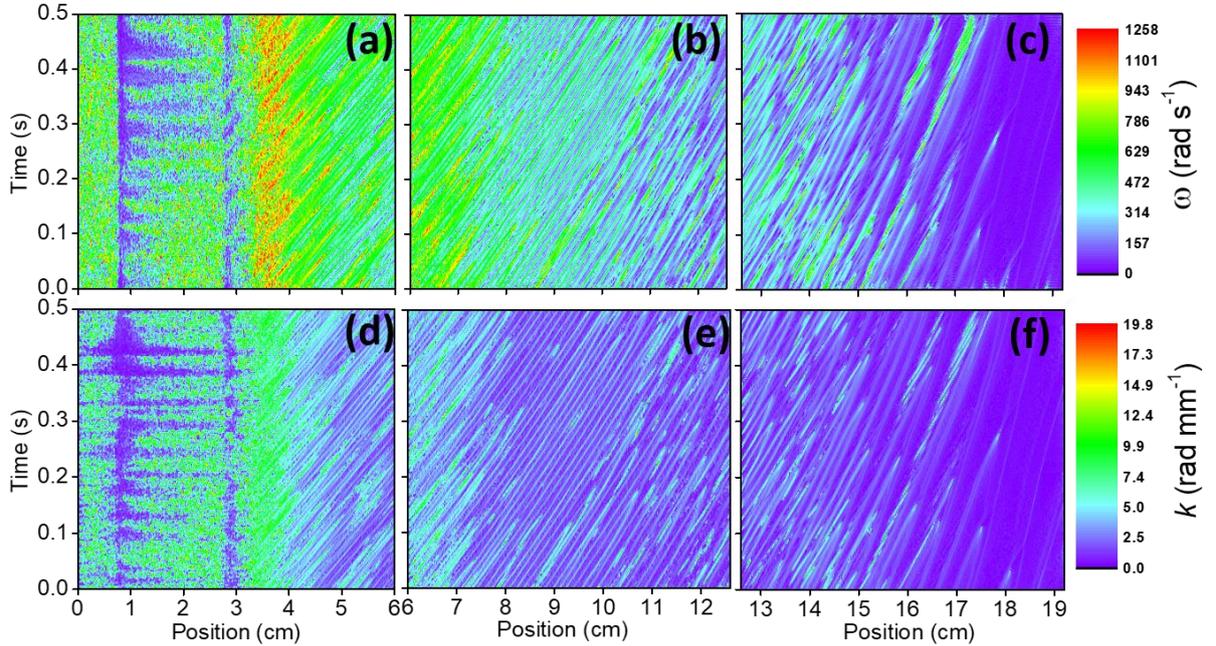

FIG. 4. (a), (b) and (c) represents the spatiotemporal evolution of frequency for the region A1, A2 and A3, respectively. The color bar represents the absolute values of frequencies in $rad\ s^{-1}$. (d), (e) and (f) represents the wave number variation along the region of interest of A1, A2 and A3, respectively. The color bar depicts the wave number values in units of $rad\ mm^{-1}$. The zero of the position axis refers to the centre of the dust void.

$$f(t) = \frac{1}{2\pi}\frac{\partial \phi(r,t)}{\partial t} \qquad (2)$$

$$k = \frac{\partial \phi(r,t)}{\partial r} \qquad (3)$$

The mean frequency at a particular position in the wave field is obtained by averaging over the complete time series, $<f(t)>$ and similarly the mean wave number is found for the entire time series. The resulting instantaneous frequencies and wave numbers are plotted in a periodogram to understand the evolution in space and time. The spatiotemporal evolution of frequency and wave number for the regions A1, A2 and A3 (marked in Fig. 3) are shown in Fig. 4 and for regions B and C, see Fig. SM3 of Supplementary Material. The inclined strips show the evolution of frequency and wave number of a particular wavefront. Besides the Hilbert transform, we have also performed Fast Fourier transformation (FFT) on the time-series signals of the DDW at each position of the region of interest and both Hilbert transformation and FFT provide a very good estimation of the frequency of the DDW (see Fig. SM4 of Supplementary Material).

From Fig. 4 (a)-(c) it is clearly seen that the frequency of the DDW is not constant for region A (along the axial direction) and has a gradient from the void edge to the bottom of the cloud. The value decreases roughly by a factor of 6 at the bottom. For the RoI (region A), the frequencies throughout the nanodust cloud form clusters of almost same frequencies, as reported previously [60,63]. Those clusters can be seen from Fig. 4 (a)-(c) as regions of almost same colour (frequency) in the periodogram and successively clusters of lower frequencies appear as one move from the void to the bottom. There can be also seen sudden changes in frequencies and wave numbers along the inclined strips throughout the whole region A which coincides with the spatiotemporal point of wave merging. As the wave travels towards the bottom, the inclined strip of frequency or wave number also broadens similarly as the case of intensity periodograms indicating the growth of non-linearity in the DDW. For region B and C, the decrease in frequencies and wave number along its propagation is not smaller compared to region A.

## IV. DISCUSSIONS

The dust void is formed just near the powered electrode as a result of higher ionization causing an outward ion drag force and an inward electric field force on the negatively charged nanodust particles. The effect of dust pressure along with these two forces, also plays an important role in the formation of dust voids in nanodusty plasmas as discussed in earlier reports [72,80,81]. The void edge is not oscillating as can be seen from the periodograms of intensity profiles of the DDW and hence the self-excited DDW is believed to be the outcome of ion-dust streaming instability [82] and not due to void oscillations. The wave appears in the cloud a few cm away (2.5 cm for region A, 1.4 cm for region B and 2 cm for the region C) from the void edge. These are very clear from the periodograms. The slopes in the periodograms are found to be lowest in region A1 and highest in A3, suggesting that the phase velocity is decreasing from region A1 (~ 9 $cm\ s^{-1}$) to region A3 (~ 6 $cm\ s^{-1}$). To explain these behaviour of the DDW, we attempted to find out the global properties of the wave such as frequencies and wave numbers by averaging the instantaneous values over a time duration of 0.5 s. The spatial variation of these time averaged frequency and wave number values are then plotted for the three regions A, B and C along their propagation direction, and are shown in Fig. 5. The highest values of frequency and wave number is almost similar for region A and B. For region C, it is comparatively smaller. The $\omega$ then decreases almost 20 times to a value of 30 $rad\ s^{-1}$ in region A. In region B and C, we observe a small decrease of frequency around one third to a value of ~ 200 $rad\ s^{-1}$. The $\omega$ and $k$ varies identically as the wave propagates in region B and C indicating an acoustic type nature of the wave. On the other hand, in region A, though initially $\omega$ and $k$ vary identically upto a distance of 5 cm from the centre of the void, beyond this distance they differ in their rate of change and finally saturate at farther distances (~ 14 cm onwards) as can be seen in Fig. 5 (a). In this saturated regime, both the quantities $\omega$ and $k$ again varies identically. In all the three regions, slight increase in $\omega$ and $k$ values at the plasma edge have been noticed. The spatial variation of phase velocities in the three regions are obtained from the ratio of time averaged $\omega$ and $k$ for each position, as shown in Fig. 6 for the three specified regions. It is seen that the phase velocity is almost constant and same ($\sim 6\ cm\ s^{-1}$) for region B and C as the frequency and wave number variation is identical. In region A, the phase velocity is $\sim 5.4\ cm\ s^{-1}$ at the beginning of the DDW and then gradually decreased towards the end with fluctuations of some higher values in between. These fluctuations of higher phase velocity are attributed to the events of wave merging or splitting where sudden increase or decrease of $\omega$ and $k$ is found, as seen in Fig. 4 (a)-(f). At the end of region A, the phase velocity decreases to $\sim 3\ cm\ s^{-1}$.

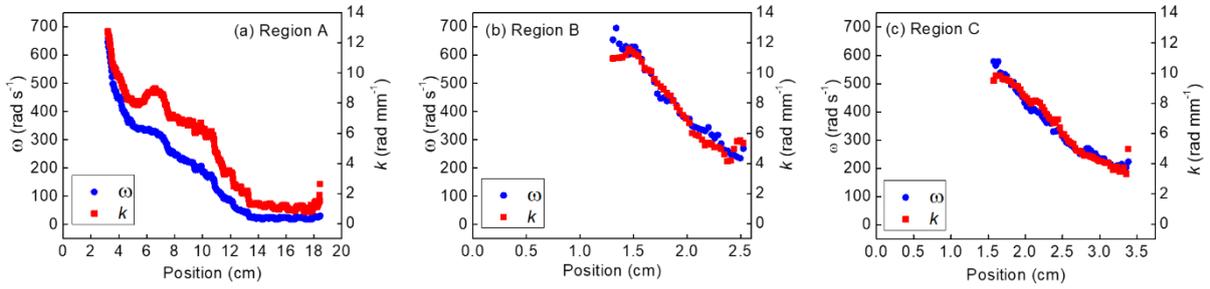

FIG. 5. Spatial variation of $\omega$ and $k$ along the three regions A, B and C.

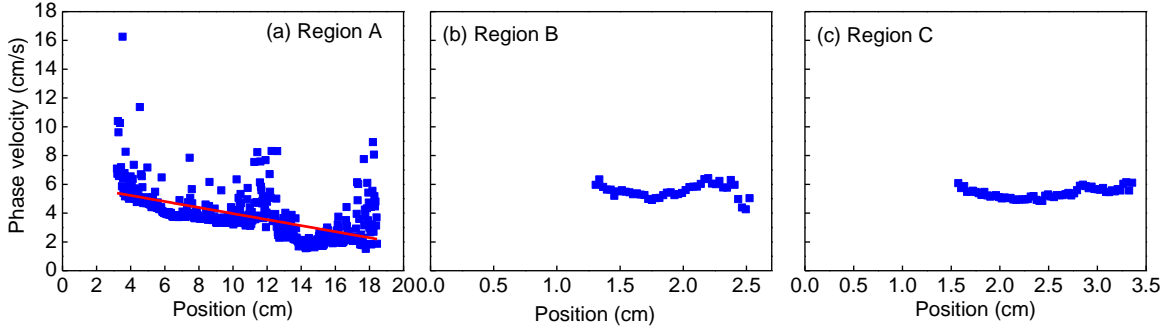

FIG. 6. Phase velocity of the wave as a function of axial positions for the wave along (a) region A, (b) region B and (c) region C. The solid (red) line is a linear fit to the plot.

We also compared the wave spectra in the three regions A, B and C. For this, a 2D Fourier transform of the periodogram of intensity in the three regions are performed. For region A, the intensity periodogram is combined for all three sub-regions A1, A2 and A3 and the corresponding wave spectra is plotted. The wave spectra reveals the map of DDW energy distribution in wave number – frequency space [39,44,54]. The wave spectra for region A, B and C are shown in Fig. 7 (a), (b) and (c), respectively. Interestingly, the wave spectra for region A have come out to be very different than the other two regions. It can be noted that, there is no or very small wave activity beyond $900\ rad\ s^{-1}$ for region A. For region B and C, beyond $\sim 500 - 600\ rad\ s^{-1}$, there is no wave activity at all. Hence it is confirmed that the frame rate of recording (400 fps) is sufficient enough to avoid aliasing in the frequency domain. In all the regions, the wave frequency is found to be almost proportional to wave number meaning an acoustic like dispersion of the DDW. It is also noted that the wave energy is unequally distributed among different frequencies and found to be higher and mostly concentrated in the low frequency regime (large wavelength) within $k = (0.1 - 2.5)\ rad\ mm^{-1}$ in region A. The maximum spectral density of the wave spectrum is approximately at $60\ rad\ s^{-1}$ (See Fig. SM5 of Supplementary Material). Recently, a possibility of a non-linear coupling mechanism is pointed out that could exchange energy from the highest wave energy region to other low energy regions [44]. It is also found that the wave energy in region A is significantly higher than the other two regions B and C which means that the dominant direction of wave propagation is in the axial direction only i.e., along region A. The slope of the dispersion relation provides the dust acoustic wave speed. In region A, the wave spectra are quite broad unlike region B and C. A broad wave spectrum means there exist variation of phase velocity along the propagation of the DDW, which is also observed earlier from the spatial phase velocity profile and from periodogram of intensity, in region A. The phase velocities obtained from the slope of the dispersion relation for the three regions are found to be same that are obtained from Hilbert transformation.

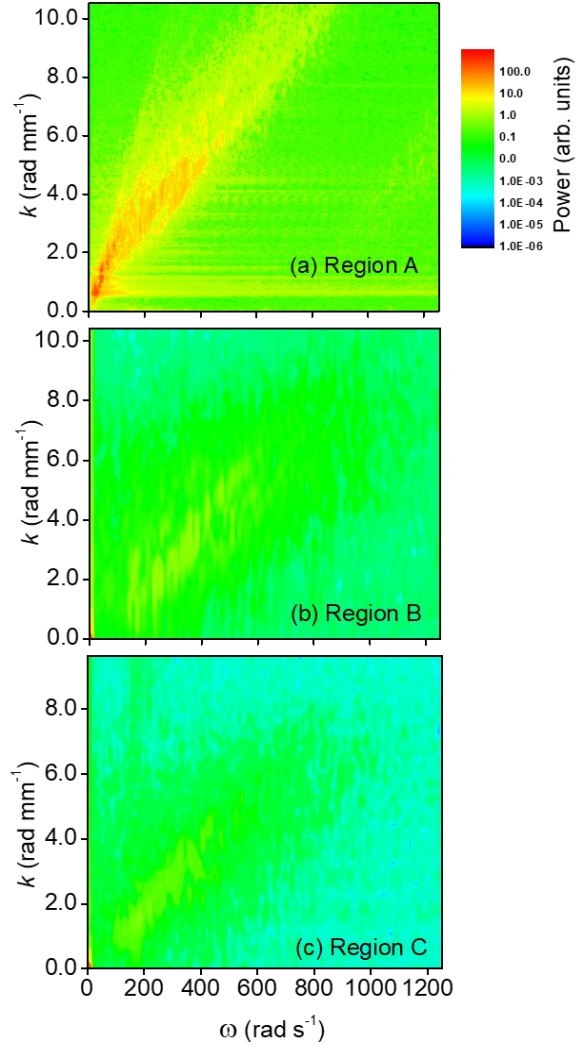

FIG. 7. Wave spectra calculated by performing 2D Fast Fourier Transformation of the periodogram for (a) region A (combining all 3 sub-regions: A1, A2 and A3), (b) region B and (c) region C.

A mismatch in the phase velocity values obtained from periodogram of intensity and from $\omega/k$ is observed as the linear fitting of spatial phase velocity profile gives a temporally averaged value of phase velocity. Whereas, in periodogram of intensity, we select only a small section of the wave front which belongs to a smaller temporal window giving us a different value compared to the global one. In general the phase velocity of a DDW in nanodusty plasma is comparatively higher than that of micron dusty plasma, as seen in earlier works [39,40]. But as the dust acoustic phase velocity, $C_D$ is directly proportional to average particle charge, there might be a decrease of phase velocity due to the Havnes effect [39,70] because of which the average dust particle charge gets reduced to a few tens of electronic charge in a dense nanodusty plasma environment. The phase velocity is defined as, $C_D = \sqrt{\frac{k_B T_i}{m_d} \frac{n_{d0}}{n_{i0}} Z_d^2}$. Here $k_B$ is Boltzmann constant, $T_i$ is the ion temperature, $m_d$ is the nanodust mass, $n_{i0}$ is the ion density and $Z_d$ is the average dust charge on each nanodust particle. Since our nanodust cloud has a gradient of dust density, from $10^{13} \, m^{-3}$ near the dust void (where the wave originates), to $10^{12} \, m^{-3}$ at the bottom of the nanodust cloud, the average nanodust particle charge may also have a gradient along the vertical direction. The estimation of dust charge in a dense nanodust cloud is very challenging as electrostatic probes cannot be used to estimate plasma parameters in the presence of nanodust particle. As the nanodust cloud is highly dense and Havnes effect comes into play and OML theory cannot be used to estimate the charge. The nanodust particle charge is heavily reduced in such an environment where the Havnes parameter is higher than unity. The Havnes parameter is defined as $P = \frac{n_{d0} Z_d}{n_{i0}} = \frac{4\pi\epsilon_0 a k_B T_e}{e^2} \frac{n_{d0}}{n_{i0}}$, where $\epsilon_0$ is the permittivity of free space, $e$ is the electronic charge, $T_e$ is the electron temperature and $a$ is the particle radius. Therefore, we have used DDW-diagnostic technique to estimate the dust particle charge, $Z_d$ using the dispersion relation of the wave which has been proven to be very effective in dense nanodusty plasma sustaining self-excited DDW [39,71].

We have used the spatially resolved frequency and wave number values obtained from Hilbert transformation to plot the dispersion relation for the three regions. The experimental dispersion relations are compared with the following theoretical model [39,42] (which includes ion streaming effect as well as collision with neutrals) with $Z_d$ and $u_{i0}$ (ion streaming velocity) as fitting parameters, and are shown in Fig. 8 (a), (b) and (c) for region A, B and C, respectively.

$$1 + \frac{1}{k^2 \lambda_{De}^2} - \frac{\omega_{pi}^2}{(\omega - k u_{i0})(\omega - k u_{i0} + i\nu_i) - k^2 v_{Ti}^2} - \frac{\omega_{pd}^2}{\omega(\omega + i\nu_{dn}) - k^2 v_{Td}^2} = 0 \qquad (4)$$

Here, $\lambda_{De} = \sqrt{\epsilon_0 k_B T_e/n_{e0}e^2}$ is the electron Debye length, $n_{e0}$ is the electron density, $k_B$ is the Boltzmann constant, $\omega_{pi} = \sqrt{n_{i0}e^2/\varepsilon_0 m_i}$ is the ion plasma frequency, $m_i$ is the ion mass, $\omega = 2\pi f$ is the angular wave frequency, $u_{i0} = \mu E$, $\mu = e/m_i \nu_{in}$ is the ion mobility, $\nu_{in} = v_{Ti} n_n \sigma_{in}$ is the ion-neutral collision frequency, $v_{Ti} = \sqrt{k_B T_i/m_i}$ is the ion thermal speed, $n_n$ is the density of neutral atoms, $\sigma_{in}$ is the ion-neutral collision cross-section, $E$ is the electric field, $\nu_i = \nu_{in} + \nu_{id}$ is the ion collision frequency, $\nu_{id}$ is the ion-dust collision frequency, $\omega_{pd} = \sqrt{Z_d^2 e^2 n_{d0}/\varepsilon_0 m_d}$ is the dust plasma frequency, $\nu_{dn} = 4\pi a^2 n_n m_n v_{Tn}/3m_d$, is the dust-neutral collision frequency, $m_n$ and $v_{Tn}$ are the mass and thermal velocity of the neutrals, respectively [42,83], $v_{Td} = \sqrt{k_B T_d/m_d}$ is the dust thermal velocity, and $T_d$ is the dust temperature. Here we have considered electron depleted dusty plasma for which the quasineutrality condition reduces to $Z_d n_d = n_i$. The number density varies in the range ~ $10^{13} - 10^{12}\ m^{-3}$ for region A. For region B and C, the number density does not vary that much and is of the order of ~ $10^{13}\ m^{-3}$. In region A, two theoretical curves are drawn for two sets of parameters, one associated with particle density $n_{d1} \sim 1.36 \times 10^{13}\ m^{-3}$ near the void (in region A1, where wave activity just appears at a distance of ~ $2.5\ cm$ from the centre of the dust void) and the other one associated with particle density $n_{d2} \sim 1.97 \times 10^{12}\ m^{-3}$ in the lower portion of the cloud (the end point of region A3). These dust density values are obtained from Fig. 2. In Fig. 8 (a), the blue and red curves represent the theoretical values of real $k$ versus $\omega$ obtained by using Eq. (4) for the above mentioned two sets of parameters related to $n_{d1}$ and $n_{d2}$, respectively. The parameters used in the theoretical calculations are $T_i = T_d = 0.025\ eV$, $v_{Ti} = 244\ ms^{-1}$, $v_{Td}$ is negligibly small (~ $10^{-3}\ ms^{-1}$) compared to the wave phase velocity and $v_{Ti}$, $\nu_i \sim \nu_{in} = 1 \times 10^6 s^{-1}$, $\nu_{id} << \nu_{in}$ (calculated by using the model given by *Barnes* et al. [84]), $\nu_{dn} = 268.5\ s^{-1}$, $T_e = 8.1\ eV$ (measured by using Langmuir probe in absence of nanodust cloud). In Fig. 8 (a), it is seen that the two theoretical curves fit well with the experimental values for $Z_d = 113$, $u_{io} = 248\ ms^{-1}$ (blue curve), and $Z_d = 55$, $u_{i0} = 255\ ms^{-1}$ (red curve). The corresponding electric field can be estimated using $E = u_{i0}/\mu$. The electric field value is found to be $E = 1.04\ V/cm$ near the dust void for $u_{io} = 248\ ms^{-1}$ and $E = 1.07\ V/cm$ at the bottom of the nanodust cloud. Therefore, it is found that throughout the nanodust cloud there exists a finite electric field because of which the ions drift from the dust void to the bottom. The ratio of ion thermal to ion drift velocity is $v_{Ti}/u_{i0} \sim 1$ for the entire nanodust cloud. For regions B and C, all other parameters are similar with region A except the dust density. The dust densities in these two regions are also of the order of $10^{13}\ m^{-3}$ and hence from the fitting of wave dispersion relation with the theory, the particle charge is found to be 135 for both the regions with ion streaming velocity of $246\ ms^{-1}$.

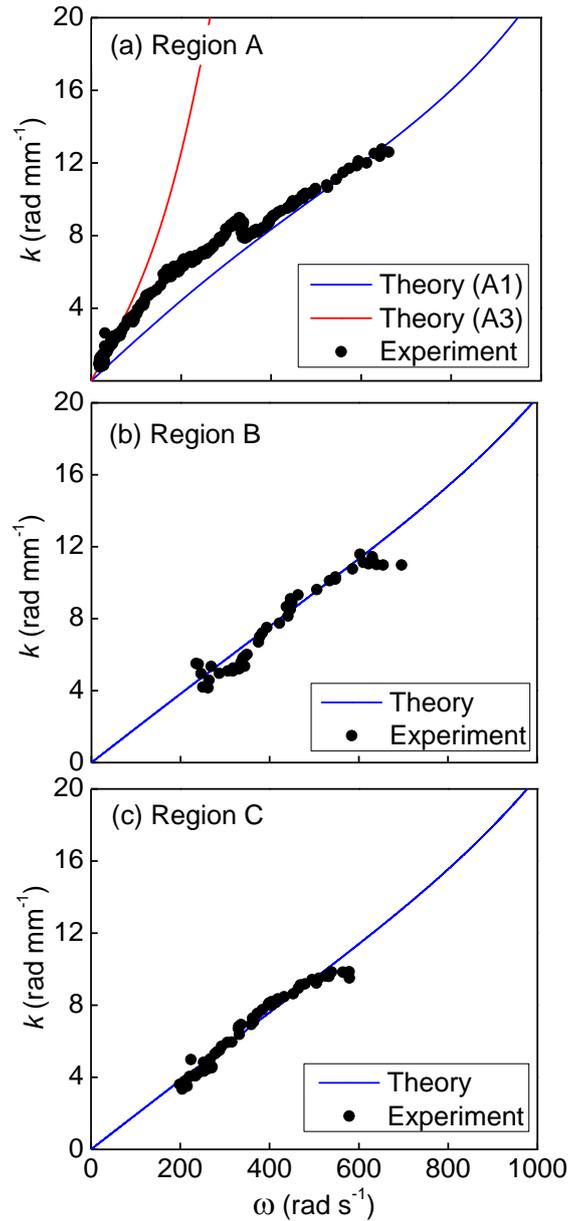

FIG. 8. Theoretical fitting of the experimentally obtained dispersion relations for the three regions A, B and C. For region A, two theoretical curves are fitted for two extreme points, $n_{d1}$ in sub-region A1 and $n_{d2}$ in sub-region A3. The solid lines represent the theoretical dispersion relations.

To estimate the spatial variation of average charge of nanodust particles along the vertical direction, we have segmented the experimental dispersion relation of Fig. 8 (a) into segments of small regions with a step size of 5 mm. For each of these small sections, theoretical curve fitting is done and $Z_d$ is estimated as described earlier. A plot of $Z_d$ versus vertical position is shown in Fig. 9. The zero of the position axis corresponds to the void centre. The dust charge is found to monotonically decrease from $Z_d = 113$ to $Z_d = 55$ within a distance of around 13 cm. From 13 cm onwards, a constant dust charge of $Z_d = 55$ has been observed. This might be due to the fact that from 13 cm onwards both the wave frequency and wave number also saturates as seen in Fig. 5 (a). After having a quantitative value of dust charge with positions, we have estimated the ion density from the quasineutral condition, along the axial direction. The ion density at a distance of 2.5 cm away from the centre of dust void (the location along region A where DDW just starts) is found to be $1.54 \times 10^{15}\ m^{-3}$ which decreases to $1.08 \times 10^{14}\ m^{-3}$ at the bottom of the dust cloud. The corresponding Havnes parameter is found to increase from $P = 15$ near the dust void to $P = 32$ at the bottom of the cloud. Now using the average dust charge and ion density estimated from the DDW-diagnostic technique, the dust acoustic wave velocity is found to be $5.8\ cm\ s^{-1}$ near the dust void and $4.1\ cm\ s^{-1}$ at the bottom. These values of DDW speed are very close to that we obtained from the experiment using time-resolved Hilbert transform (Fig. 6 (a)). Therefore, we believe that the cause of smaller phase velocity of the DDW is the Havnes effect. The Havnes parameter, $P$ is found to be much higher than unity meaning a serious reduction of dust charge. As the Havnes parameter increases towards the bottom of the cloud, the wave velocity also decreases due to the heavy reduction in dust charge. For region B and C, the Havnes parameter remains fixed ($P = 13$) and does not vary much with distance and hence the wave travels with a constant speed along the radial and oblique direction.

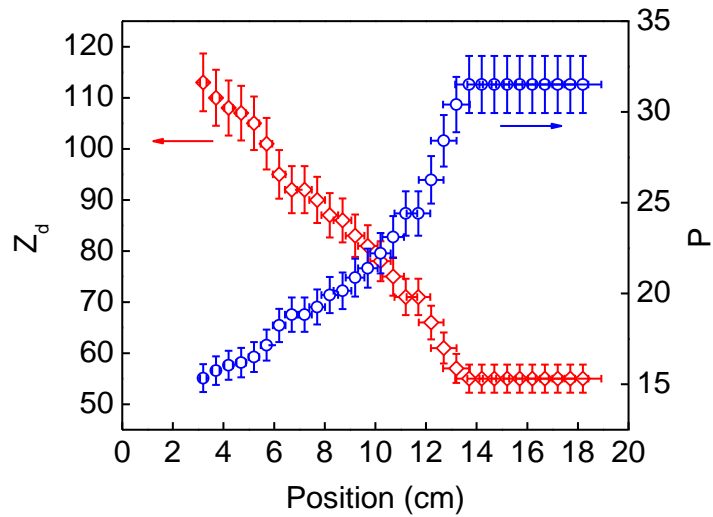

FIG. 9. Spatial variation of average charge, $Z_d$ on each nanodust particle and Havnes parameter, $P$ along the axial direction i.e., region A.

With these understanding of the DDW we attempted to explore the formation of dust void mechanism by estimating various forces quantitatively. The dust void forms as a result of a balance between outward ion drag force and inward electric field force exerted on the dust particles. Along with these two forces a dense dust cloud can exert a sufficient amount of dust pressure also, as discussed earlier [72,80,81]. In the 2D central plane, the dust void dimension is $16 \times 7\ mm^2$ in our experiment. We estimate the ion drag force exerted on the dust particles (not on the edge of the dust void but on the location where the DDW just starts to appear i.e., 2.5 cm away from the centre of the dust void) using Barnes model [84] in a similar way as discussed in *Bailung* et al. [24]. The ion drag force is calculated to be $F_{ion}^{axial} = 2.64 \times 10^{-15}\ N$ along the axial direction and $F_{ion}^{radial} = 3.95 \times 10^{-15}\ N$ along radial direction. The electric field force is $1.88 \times 10^{-15}\ N$ and $2.25 \times 10^{-15}\ N$ along the axial and radial direction, respectively. These two opposing forces are of the same order which can easily counter each other and form a dust void. The values of the forces may vary minimally at the edge of the dust void due to the difference of dust charge and number density but we believe the order will remain same as estimated here. The dust pressure is also estimated using the expression given in [72,80] and found out to be $6.01 \times 10^{-14}\ N$ along axial and $1.56 \times 10^{-14}\ N$ along the radial direction.

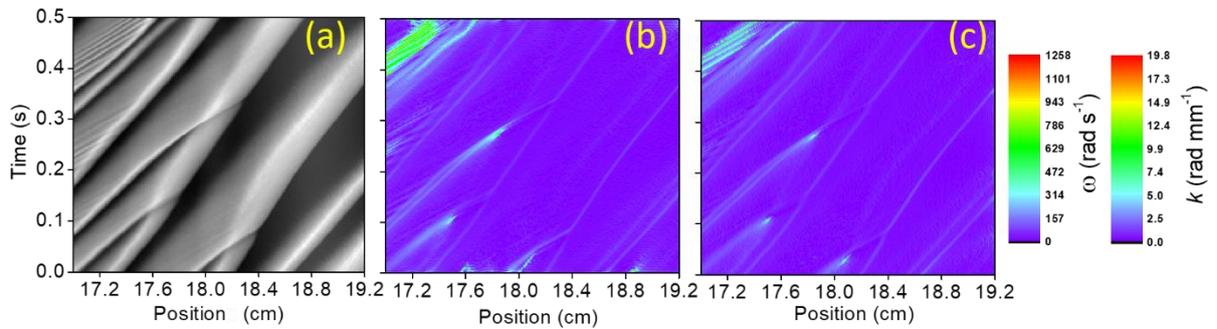

FIG. 10. (a), (b) and (c) are enlarged sections of Fig. 3 (f), Fig. 6 (c) and (f), respectively. (a) shows the spatiotemporal evolution of wave fronts during wavefront merging events. (b) and (c) represent the spatiotemporal evolution of frequency and wave number, respectively. The corresponding colour bars are shown in the right.

The self-excited DDW is also seen to sustain various topological defects corresponding to wave merging phenomena. Wavefront merging occurs when the wave fronts travel with different velocities. An enlarged section of the intensity periodogram (Fig. 3 (f)) of the DDW is shown in Fig. 10 (a). The spatiotemporal evolution of frequency and wave number, corresponding to Fig. 10 (a) are shown in Fig. 10 (b) and (c), respectively. In our self-excited

DDW, the phase velocity gradually decreases along the axial direction. From Fig. 10 (a), it has been observed that the wave fronts change their velocity before and after merging. From the periodogram of intensity, it is observed that the velocity of the resultant wavefront is smaller than that of the individual wavefronts. The amplitude of the resultant wavefront is also observed to increase after merging. One very peculiar characteristics of the merging phenomena that has been observed from the spatiotemporal evolution of wave frequency and wave number, shown in Fig. 10 (a) and (c) respectively, is that just before the merging event, both the frequency and wave number increase to a sufficiently high value and after the event, they gets reduced to a lower value (smaller than that of individual wavefronts). The frequency enhancement is found to be almost one order higher before every wave merging event.

## V. CONCLUSION

A self-excited DDW propagating in a vertically extended nanodusty plasma ($\sim 19\ cm$) is studied in detail in terms of their propagation throughout the nanodust cloud along three directions (axial, radial and oblique). Time-resolved Hilbert transformation is used to estimate the spatiotemporal evolution of wave frequency and wave number and finally spatial variation of phase velocity along all three directions are obtained. Fast Fourier transformation is utilized to study the wave spectra in the three regions. The wave is seen to be dominantly propagating in the axial direction as the wave energy is mostly confined in the same direction. The DDW evolves from a coherent state near the dust void to a turbulent one at the bottom indicating about the onset of non-linearity in the DDW as it propagates. This is more prominent in the axial direction. The growth rate of the wave amplitude is found to be $0.3\ cm^{-1}$. Along the radial and oblique directions, the turbulent nature and non-linearity appear to be smaller as compared to that in the axial direction. Experimental dispersion relations are obtained for the three directions and compared with a theoretical model. From the comparison, nanodust particle charge and ion density of the plasma in presence of the dust cloud are obtained. A finite electric field is also found to be present throughout the plasma which is believed to be the cause of DDW excitation. The Havnes parameter is found to be quite higher than unity ($P = 15 - 32$) because of which the particle charge is strongly reduced and DDW is slowed down as it propagates in the downward direction away from the dust void. Our estimates for the electric field value, ion streaming velocity and dust particle charge agree well with previous similar works. A quantitative analysis of the dust void formation is also presented and a few unique characteristics of wavefront merging phenomena are explored. A detail analysis of the development of non-linearity and the evolution of the structure of wavefronts as the DDW travels to a large distance are under progress.

## ACKNOWLEDGEMENTS

BC would like to thank DST, Govt. of India, for financial support under INSPIRE fellowship. One of the authors (SKS) thanks DST, Govt. of India for supporting the work under the Women Scientist (WOS-A) project scheme. The Authors thank Benjamin Tadsen for fruitful suggestions regarding the Hilbert transformation.

# Supplementary Material

**Spatiotemporal evolution of a broad-spectrum dust density wave under strong Havnes effect in a vertically extended nanodusty plasma**

Bidyut Chutia[1], T. Deka[1,2], Y. Bailung[1,3], D. Sharma[1,4], S. K. Sharma[1,5*] and H. Bailung[1]

[1]*Physical Sciences Division, Institute of Advanced Study in Science and Technology (IASST), Paschim Boragaon, Guwahati-781035, Assam, India*

[2]Presently at *Department of Physics, Sipajhar College, Mangaldoi-784145, Assam, India*

[3]Presently at *Department of Physics, Goalpara College, Goalpara-783101, Assam, India*

[4]*Academy of Scientific and Innovative Research (AcSIR), Ghaziabad-201002, India*

[5]Presently at *Department of Physics, Assam Don Bosco University, Tapesia Gardens, Sonapur-782402, Assam, India*

*sumita_sharma82@yahoo.com

The Supplementary Material consists of this single PDF file only.

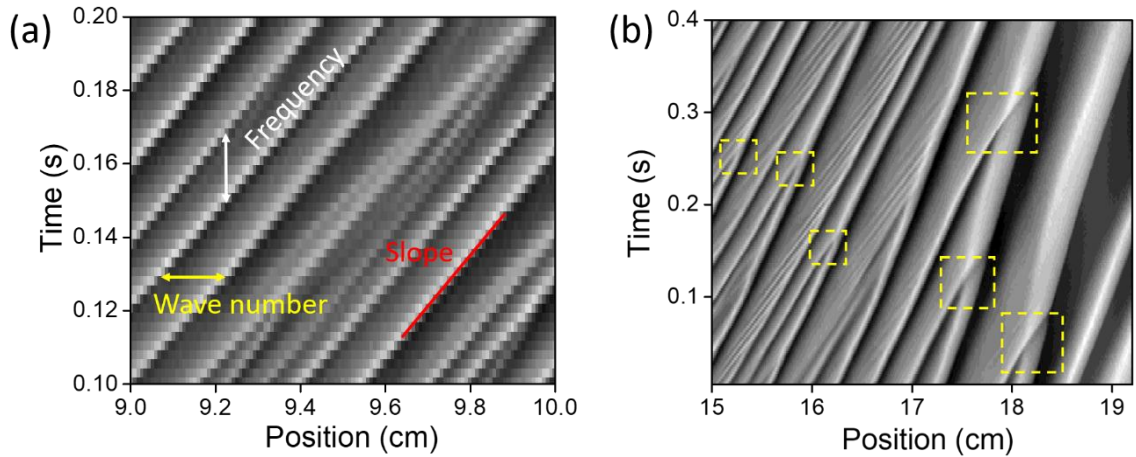

Fig. SM1. (a) A small section of region A2 is shown here to elaborate how frequency and wave number can be extracted from periodograms. The white arrow represents the temporal difference between two wave fronts and the inverse of this difference will give the frequency in Hz. The yellow arrow represents the spatial difference between two successive wave fronts and inverse of this spatial difference will give the wave number. Inverse of slope (red line) of the wave front yields the wave velocity. (b) A section from region A3 is presented to show the merging of wave fronts as the wave propagates. The yellow squares show the various merging events.

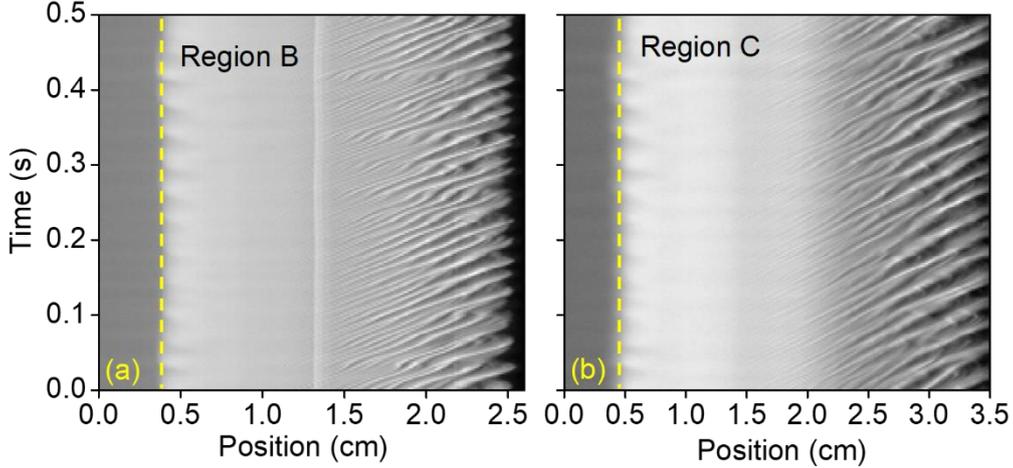

Fig. SM2. Periodogram for (a) region B and (b) region C. The zero of the position axes refers to the centre of the dust void in both the periodograms. (b) shares the same y-axis with (a). The dashed (yellow) lines represent the void edge.

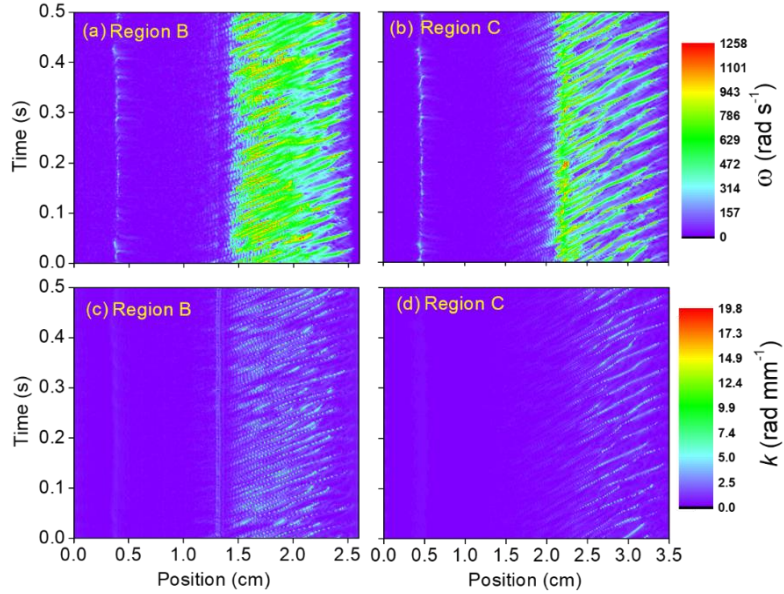

Fig. SM3. (a) and (b) represents the spatiotemporal evolution of frequency for the region B and C of Fig. 3 (a), respectively. The color bar represents the absolute values of frequencies in $Hz$. (c) and (d) represents the wave number variation along the region of interest of B and C, respectively. The color bar depicts the wave number values in units of $rad\ mm^{-1}$. Subfigure (a) and (b) shares the same position axes as subfigure (c) and (d), respectively. The zero of the position axis corresponds to the centre of the dust void.

The wave activity in region B and C is confined to only within 1 cm and 1.5 cm, respectively. The inclined strips of frequencies in region C are seen to be segmented meaning there is a large variation of frequencies of a single wavefront in the spatiotemporal scenario as the DDW is propagating to a larger distance and hence growing to a larger non-linearity and forming various defects points compared to B.

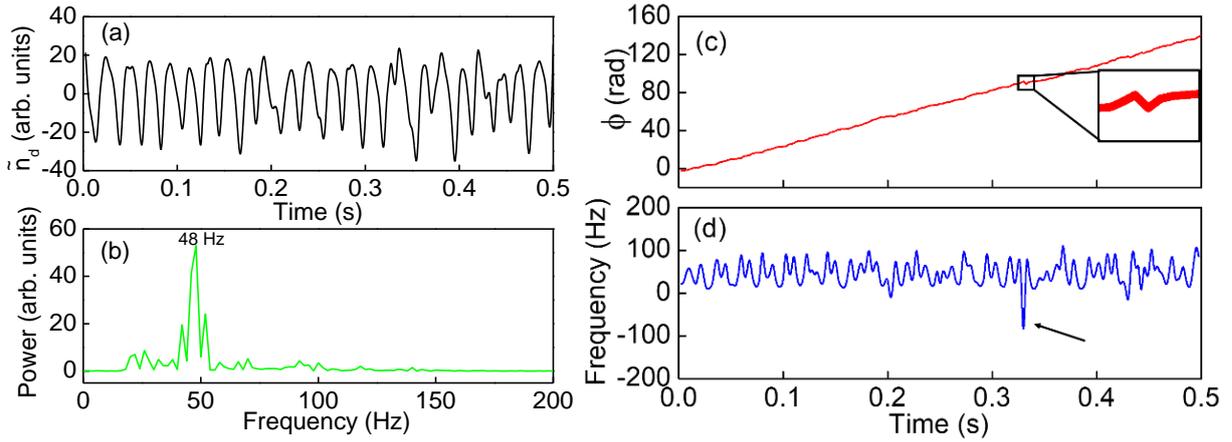

Fig. SM4. (a) Typical time-series signal of intensity fluctuation (average) of the DDW at a particular position for a duration of 0.5 s. (b) Fourier transformation of the time-series signal. (c) and (d) are the phase and frequency spectrum obtained by using time-resolved Hilbert transformation for the same signal in (a). (c) and (d) share the same x-axis. It is observed that the average of the frequency spectrum of (d) is almost equal to the dominant frequency component obtained from Fourier transformation which is ~ 48 Hz. In the inset of (c), a defect is shown which comes up as a sudden drop in phase profile. In general, the phase is a monotonically increasing function of time but it might change in locations where topological defects are present. In (d), corresponding to the defect, a drop in frequency value (black arrow) is shown. It is seen that the frequency values are oscillating in time which is a manifestation of non-linearity of the wave.

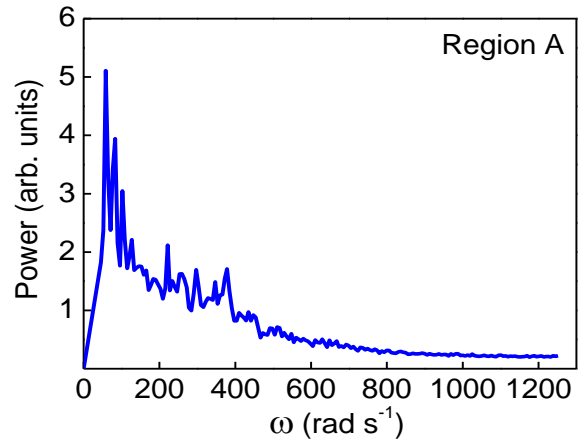

Fig. SM5. Spectral density distribution over the entire frequency range for region A.